\documentclass{article}
\usepackage{graphicx,amsmath}
\newtheorem{theorem}{Theorem}

\newtheorem{conjecture}{Conjecture}
\begin{document}

\title{Price's law,
mass inflation, and strong cosmic censorship}

\author{\it Mihalis Dafermos\\
\rm Department of Mathematics, MIT
\\ 77 Mass Ave Cambridge, MA 02139 U.S.A.\\
E-mail:dafermos@math.mit.edu}

\maketitle

\abstract{%
Two aspects of the widely accepted heuristic picture of the final
state of gravitational collapse are the so-called Price law tails,
describing the asymptotics of the exterior region of the black hole
that forms, and Israel-Poisson's mass 
inflation scenario, describing the internal
structure of the black hole. (The latter scenario, if valid, would
indicate in particular that the maximal development of initial
data is extendible as a $C^0$ metric, putting into question
the validity of Penrose's strong cosmic censorship conjecture.) In
this talk, I shall discuss a series of rigorous results 
proving both Price's law and the mass inflation scenario in an appropriate 
spherically symmetric setting. The proof of Price's law is joint work
with I.~Rodnianski.}

\section{The problem at hand}  \label{intro}

A central physical problem in general relativity is the study 
of the collapse of isolated self-gravitating systems. 
This process gives rise to spectacular predictions: the possible formation
of black holes, naked singularities and Cauchy horizons. These predictions
are given precise mathematical formulations in the celebrated 
weak and
strong cosmic censorship conjectures of Penrose~\cite{rp:gcsts, rp:sst}.
Ultimately, resolution of these conjectures, without any additional
symmetry assumptions, constitutes perhaps the central goal of 
mathematical relativity.

In reaching this goal, it is clear that there are 
several major obstacles that will have to 
be overcome. I believe that the first
is the complete understanding of a ``realistic'' 
spherically symmetric formulation, i.e.~a rigorous 
analysis of the global initial value problem
for an appropriate Einstein-matter system, with spherically 
symmetric initial data, including, in particular, a proof (or disproof) of 
both cosmic censorship conjectures.
With the word ``appropriate'', it is required that the dynamical
degrees of freedom of the gravitational field, together with the centrifugal
force of angular momentum--both absent in spherical symmetry!--be somehow, 
nevertheless, modeled. In the case where the effects of angular momentum
are ignored, this problem was resolved by D.~Christodoulou~in a
series of papers~\cite{chr:sgsf, chr:fbh, chr:ins}.
In this talk, I will describe my own contributions to this
program~\cite{md:si, md:cbh, mi:mazi} where the repulsive
effects of angular momentum are modeled with charge. As we shall see,
the implications for strong cosmic censorship are completely different.
My results are motivated by previous heuristic and numerical work.

Part of the above program~\cite{mi:mazi} involves proving
the so-called Price's law. This is joint work with Igor Rodnianski. 
Although our motivation was the role Price's law plays in
the mass inflation scenario, for which charge must be present, 
our proof of Price's law applies in addition in the absence of charge, where
it has independent interest. Moreover, our results also
apply to the linearized problem of the wave equation on a Schwarzschild
or Reissner-Nordstr\"om background. In this linear case, weaker results can
also be deduced in the case where the scalar field is not
spherically symmetric.

\section{The Cosmic Censorship Conjectures}

For appropriate coupled Einstein-matter systems,
one can associate to initial data
a unique maximal globally hyperbolic spacetime $(M,g)$,
the so-called \emph{maximal development}.
The fundamental mathematical and physical 
questions about gravitational collapse refer then to the global geometry
of this spacetime. In spherical symmetry, precise geometric properties
can be read off a diagram
 which we can associate\footnote{The quotient manifold $\mathcal{Q}=M/SO(3)$
of the group action can be be endowed with a $1+1$ dimensional Lorentzian
metric; a Penrose diagram of $M$ is then just the image
of a conformal representation of $\mathcal{Q}$ into a bounded
domain
of $1+1$ dimensional Minkowksi space. 
Thus the causal relation of two points can be read off the diagram,
as if these points were in $1+1$ dimensional Minkowski space.}
to $M$, its so-called Penrose diagram.

The Penrose diagram of the 
\emph{classical} conjectured picture of generic gravitational collapse
is depicted below:
\[
\includegraphics{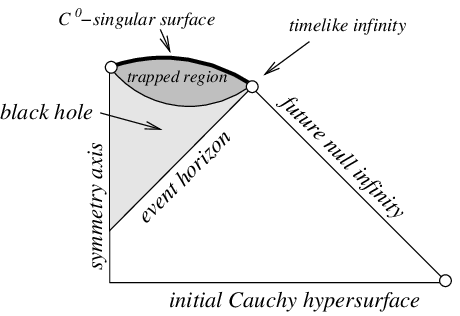}
\]
This ``conjectured'' diagram, in particular, encodes\footnote{We emphasize
again that the diagrams have precise meanings and can be treated at
the same level as formulas.} the statement
that the following two conjectures are true:
\vskip.3pc
\noindent{\bf Weak cosmic censorship}
For generic initial data, the maximal development
possesses a complete future null infinity.
\vskip.3pc
\noindent{\bf Strong cosmic censorship} For generic initial data,
the maximal development is inextendible as a $C^0$ metric.
\vskip.3pc
A proper discussion of the terminology in the formulation of these
conjectures\footnote{The above formulations are 
due to Christodoulou~\cite{chr:givp}.}
is impossible here. Hopefully, even for the non-expert,
the Penrose diagrams will be sufficiently suggestive. Informally, weak 
cosmic censorship says 
that there exist observers who observe a regular past forever,
 and strong cosmic 
censorship says that observers 
reaching the end of spacetime must be destroyed.
Under the above formulations,
the 
weak and the strong cosmic censorship conjectures
are in fact independent.

The above picture of gravitational collapse
was in fact rigorously 
obtained in 1939 by Oppenheimer and Snyder~\cite{os:?}
for the simple model of a homogeneous dust\footnote{The study of this
model reduces to o.d.e.'s.}; the
significance of this work was not appreciated until the 1960's. 
Christodoulou~\cite{chr:dust}, however, 
showed that the Oppenheimer-Snyder model is unstable to small
inhomogeneities, as naked singularities form due to 
shell-crossing singularities.
In the search for a more robust model that would
not form such ``spurious'' singularities, Christodoulou was led to
the Einstein-scalar field system. 
In a series of fundamental papers~\cite{chr:sgsf,chr:bv,chr:fbh, chr:ins}, 
Christodoulou was able to show that generically,
the above conjectured Penrose diagram is correct for this
system under spherical symmetry.\footnote{Moreover,
he was also able
to explicitly produce non-generic counterexamples to weak cosmic 
censorship.}
Christodoulou's work initiatied 
the rigorous theory of nonlinear p.d.e.'s
as a tool for resolving
the physical problems
of gravitational collapse.

\section{The challenge of angular momentum}

In looking ahead to the non-spherically symmetric case, one can
unfortunately not realistically hope to extrapolate from Christodoulou's model.
The spherically-symmetric Einstein-scalar field system does not account for
effects of angular momentum. 
In the non-spherical case, the Newtonian theory indicates that
these effects will dominate after collapse has progressed.
It is precisely because the goal of the spherically symmetric
study is to give insight into the non-spherical case that it is imperative
to investigate a model that can encorporate these effects. As we shall see
below, angular momentum in general relativistic collapse in fact 
introduces phenomena that have no parallel in the 
Newtonian theory.

Fortunately for the prospect of studying this problem in
the context of spherical symmetry,
it turns out that there is a close connection 
between the repulsive mechanisms of charge and the centrifugal 
force of angular momentum; one manifestation of this is the analogy
between the conformal structure of the rotating Kerr solution of
the Einstein vacuum equations and the spherically
symmetric Reissner-Nordstr\"om solution of the Einstein-Maxwell equations:
The Penrose diagram of the latter is depicted below:
\[
\includegraphics{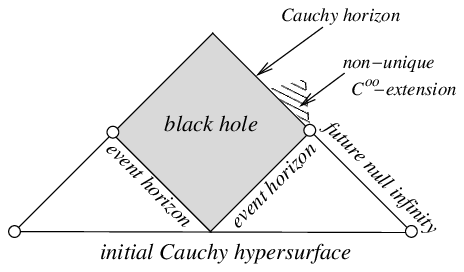}
\]
Points beyond the Cauchy horizon are not part of the maximal development
because the spacetime would there fail to be globally hyperbolic. Yet there
is no obstruction to extending beyond! It is clear, however, that such
extensions are not uniquely determined by initial data. Thus, we see
that the introduction of arbitrarily small angular momentum (or charge)
indicates that Newtonian determinism fails in a spectacular way! 

The purpose of strong cosmic censorship~is 
precisely to exclude the above phenomena, at least
generically.
In fact, S.C.C.~was originally conjectured on the basis of a geometric optics
argument that indicated that linear fields blow up on the Reissner-Nordstr\"om
Cauchy horizon.\footnote{This insight is due to Penrose. 
There is a long history of numerical, heuristic, and analytic
work on this linearized problem. See~\cite{sp:RN, ch:cchRN}.}
In a series of papers, which I will describe in this talk,
I studied strong cosmic censorship in the context 
of the Einstein-Maxwell-neutral
scalar field equations:
\[
R_{\mu\nu}-\frac{1}{2}g_{\mu\nu}R=2T_{\mu\nu},
\]
\[
g^{\mu\nu}F_{\lambda\mu;\nu}=0, F_{[\lambda\mu,\nu]}=0
\]
\[
g^{\mu\nu}\phi_{;\mu\nu}=0,
\]
\[
T_{\mu\nu}=F_{\mu\lambda}F_{\nu\rho}g^{\lambda\rho}
-\frac1{4}g_{\mu\nu}F_{\lambda\rho}F_{\sigma\tau}
g^{\lambda\sigma}g^{\rho\tau}+
\phi_{,\mu} 
\phi_{,\nu}
-\frac12g_{\mu\nu}g^{\alpha\beta}\phi_{,\alpha}\phi_{,\beta}.
\]
In spherical symmetry, the above system reduces to a hyperbolic
system of p.d.e.'s in $2$ independent variables. The Maxwell part decouples,
and contributes an $\frac{e^2}{r^4}g_{\mu\nu}$ term to the energy-momentum 
tensor, where $e$ is a constant to be called \emph{charge}.
The above system is in a sense quite
peculiar, and in particular, cannot serve as a model for gravitational
collapse from regular data with one asymptotically flat end.\footnote{This
is due to the fact that, since there is no charged matter,
the charge must be topological in origin. In particular, this implies
that trapped or antitrapped surfaces must be present on initial
Cauchy data. See the last section.}
However, it is in some sense the 
simplest hyperbolic system
which includes Reissner-Nordstr\"om as a special solution, and allows
one to study the question of the stability or instability
of the Cauchy horizon in a non-linear setting.

In~\cite{md:si}, I proved the following:
\begin{theorem}
Consider a double characteristic initial value problem,
where the outgoing characteristic is given the data
of a Reissner-Nordstr\"om event horizon with $e\ne0$, and arbitrary sufficiently
regular matching data are prescribed on the ingoing characteristic.
Then the maximal development is as depicted:
\[
\includegraphics{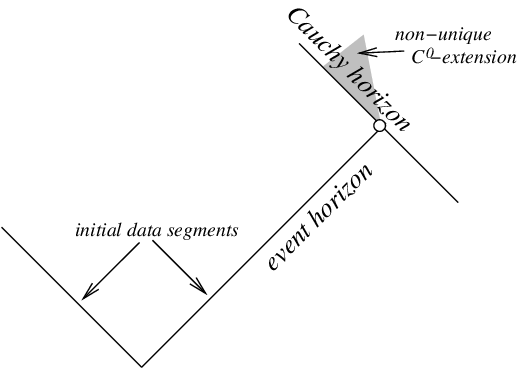}
\]
In particular, the Cauchy horizon survives and the metric can
be extended beyond it as a $C^0$ metric.
\end{theorem}
On the other hand, I also proved in~\cite{md:si}
\begin{theorem}
Assuming that some appropriate quantity, computed from the data 
at the 
point of intersection of the initial characteristic segments, is non-zero,
it follows then that the Hawking mass 
blows up identically along the Cauchy horizon.
Thus, the spacetime \emph{cannot} be extended as a $C^1$ metric.
\end{theorem}

These results were the first rigorous confirmation
of a scenario that had been originally proposed by Israel and Poisson, 
and studied in a long series of
heuristic and numerical work, e.g.~\cite{ispo:isbh, brsm:bhs,
bdim:scsbhi, lb:schs}. The picture of the above two theorems
is known as \emph{mass inflation}.

\section{The geometry of the event horizon: Price's law}

While the above theorems indeed show that the stability of the Cauchy horizon
(albeit only in the $C^0$ norm) \emph{can} in principle occur in
the non-linear theory, they do not show that it \emph{does} occur in
dynamical spacetimes arising from collapse.
For it should be clear that
the data considered above--with event horizon \emph{exactly} 
Reissner-Nordstr\"om--are 
non-generic and unphysical. One could easily speculate that
Theorems 1 and 2 are an artifice of the particular initial 
value problem posed, and that the conjectured classical
picture would be restored were the ``correct'' problem considered.
 
What are then the ``correct'' characteristic data? At the time~\cite{md:si}
was completed, this remained an open
question. There was, however, a definite conjecture, 
formulated in 1972 by Price~\cite{rp:ns}. According to the
so-called Price law, the scalar field should decay
with respect to a naturally defined advanced time coordinate along
the event horizon as $v^{-3}$.\footnote{Price's heuristics,
originally formulated with respect to Schwarzschild,
were extended to the charged case in~\cite{jb:gcc}. There has been
a long series of heuristic and numerical work confirming Price's law, see
for example~\cite{lb:sbh, bo:lt, bo:lte, mc:bhsf}.}

In~\cite{md:cbh}, I proved
\begin{theorem}
Consider a spacetime with a black hole and regular event horizon,
satisfying an appropriate upper bound formulation of Price's 
law\footnote{The condition is
$|\partial_v\phi|\le Cv^{-1-\epsilon}$ for some $\epsilon>0$. See
Theorem~\ref{timn} for the definition of the advanced time
coordinate $v$.} for the
decay of the scalar field, and $0<e<\sqrt{2m_+}r_+$, where
$m_+$ and $r_+$ denote the limiting values of the Hawking mass and
area radius, respectively, along the event horizon. 
Then the black hole contains a Cauchy horizon
over which the spacetime is extendible as a $C^0$ metric.
\end{theorem}
In addition, I proved  
\begin{theorem}
\label{katwfragma}
Assuming now a lower bound formulation of Price's law\footnote{The condition
is as follows: There exist positive constants $V$, $C_1$, $C_2$, 
and a constant $s>\frac12$,
such that $v\ge V$ implies $C_1v^{-s}\ge|\partial_v\phi|\ge C_2v^{-3s}$
along the event horizon. See Theorem~\ref{timn} for the definition
of the advanced time coordinate $v$.}, the Cauchy horizon
is $C^1$-singular, in particular the curvature blows up.
\end{theorem}
The above two theorems showed thus that the results of~\cite{md:si} were not 
an artifice of the particular initial value
problem posed, but rather, a consequence of Price's law.

Complete elucidation of the Einstein-Maxwell-neutral scalar field
picture thus reduced to \emph{proving} Price's law.  
In approaching this problem, however,
there was a major obstacle. The physical intuition that motivated 
the conjecture of power-law decay was based on heuristic arguments for
the linearized problem, i.e.~the study of the wave equation on
a fixed Schwarzschild or Reissner-Nordstr\"om background.
These arguments rest for the most part on Fourier analytic and or spectral
theoretical techniques, which do not appear sufficiently robust to carry
over to the non-linear theory. 
A new way of looking at the problem was necessary. In collaboration with Igor
Rodnianski, I developed a technique based on the interaction
of the global conformal geometry, the celebrated \emph{red-shift} effect,
and local energy conservation, to understand the behaviour of the scalar
field in the exterior. Some of the main points of this technique will be
discussed in the next section.
In~\cite{mi:mazi}, we proved
\begin{theorem}
\label{timn}
Consider spherically symmetric asymptotically flat initial data for
the Einstein-Maxwell-scalar field equations, where the scalar field
and its gradient are initially of compact support, and assume the data
contain a trapped surface. Then the maximal development of initial data
contains a domain of outer communications possessing a complete
future null infinity, as depicted:
\[
\includegraphics{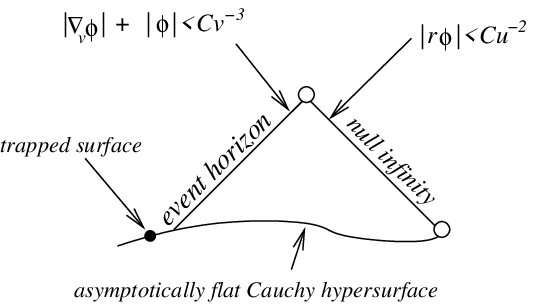}
\]
Defining a natural advanced time coordinate
$v$ on the event horizon, and a retarded time coordinate $u$ on null infinity,
then the decay rates as depicted above hold.\footnote{The 
decay rate is actually
$|\phi|+|\partial_v\phi|\le C_{\epsilon}v^{-3+\epsilon}$, 
for all $\epsilon>0$, but this
is sufficient for applying Theorem 3. In addition, it is necessary to
assume that $e<\sqrt{2m_+}r_+$,
i.e.~that the black hole is not extremal in the limit. The coordinate
$v$ is advanced time normalized so that $v=r$ on some fixed outgoing
null ray intersecting null infinity, while $u$ is retarded time normalized
so that $\partial_ur=-1$ on null infinity.} 
\end{theorem}
The upper bound formulation of Price's law thus indeed holds. 
Moreover, the proof clarifies
the physical mechanism generating decay, in particular, the origin of
the power $-3$. Our techniques 
also apply to the linearized problem of the wave equation on a fixed
Schwarzschild or Reissner-Nordstr\"om background\footnote{Machedon and
Stalker had announced a weaker version of Price's law for the linear problem,
prior to our non-linear proof.}. Since, as we noted
earlier, complete spherically symmetric 
initial data of the Einstein-Maxwell-real scalar field system necessarily
possess a trapped surface\footnote{as long as the Maxwell field does not
vanish identically}, we obtain from Theorems 3 and~\ref{timn}:

\begin{theorem}
Strong Cosmic Censorship is false for the Einstein-Maxwell-real scalar field
system under spherical symmetry.
\end{theorem}

On the other hand, in view of Theorem~\ref{katwfragma}, the 
generic $C^1$-\emph{inextendibility} of the maximal development
would be given by a positive resolution to the following:

\begin{conjecture}
For generic initial data for the problem considered in Theorem~\ref{timn},
there exist positive constants $\epsilon$, $V$, and $C$, such that along
the event horizon
$|\partial_v\phi|\ge Cv^{-9+\epsilon}$ for $v\ge V$.
\end{conjecture}

\section{Remarks on the proof of Price's law}
The proof of Price's law is rather involved, even in skeleton form. 
The main ingredients, however, when viewed separately, are quite transparent.
They can be thought of as 
relatively simple analytical manifestations or consequences of the following
robust\footnote{i.e.~features well-known from the Schwarzschild solution,
but not special to it} geometrical features of black hole spacetimes
\begin{enumerate}
\item[(A)] 
a well-defined notion of infinity,
\item[(B)]
the global conformal geometry; in particular, the existence of an affine
complete null hypersurface (the event horizon) which 
does not terminate at null infinity
and whose past is the entire domain of outer communications,
\item[(C)]
the celebrated red-shift effect.
\end{enumerate}
In this section, we will be content to make some very general 
remarks on how these features can be connected to one
another in the analysis of the system.  
For more details, the reader can consult \cite{mi:mazi}.

We first discuss (A). 
This feature is of course the least exotic of 
the above, as it is familiar
from Minkowksi space and small data 
dispersing solutions. For us, (A), together with
the special form of energy 
conservation (see below) that is valid for our system, will allow us to obtain 
\emph{a priori} uniform decay in $r$ for the scalar field, and certain
of its derivatives, especially $\partial_v(r\phi)$, which, as we shall see,
decays better.  It is this 
$r$-decay that we hope to translate into $v$-decay along the event horizon, 
and $u$-decay along null infinity.\footnote{See the footnote to 
Theorem~\ref{timn} for the definitions of the $u$ and $v$ coordinate.}

The analytic significance of (B) is that 
the event horizon carries a positive finite flux, 
arising from integrating
\[
d\left(m+\frac{e^2}{2r}\right)=
\frac12\frac{(r\partial_u\phi)^2}{\partial_ur}(1-\frac{2m}r)du
+\frac12\frac{(r\partial_v\phi)^2}{\partial_vr}(1-\frac{2m}r)dv
\]
in $u$ and $v$ throughout the domain of outer communications, and applying
Stokes' theorem.\footnote{This encorporates energy conservation in our
system. It is interesting to note that in the linearized case, the $1$-form
on the right hand side of the above equation is again closed; it is precisely
the form arising from contracting the energy momentum tensor
of the scalar field with the $\partial_t$ Killing vector field.}  
Since the event horizon
is affine complete, every so often there must be large
regions where the flux is quantitatively small:
\[
\includegraphics{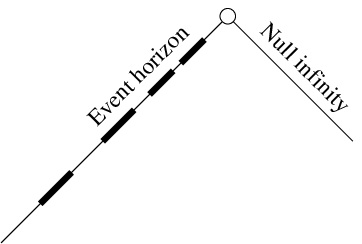}
\]
Specifically, 
by dyadically decomposing the event horizon, and then further subdividing
each dyadic interval, we will be able to choose
subintervals of size increasing as a positive power of $v$ 
such that the flux in each subinterval
decays as some power of $v$, where $v$ measures the position of the interval.
This smallness
will act in our argument 
as a catalyst for turning decay in $r$ into decay in $v$.

An efficient use of the event horizon flux in 
the above role is only possible
when one is able to transport decay from the event
horizon without much loss to other regions of spacetime, in particular, to
regions where $r$ is on the order of $v$ to some positive
power. In other words, one needs
to be able to derive good estimates in the
direction indicated in the figure below.
\[
\includegraphics{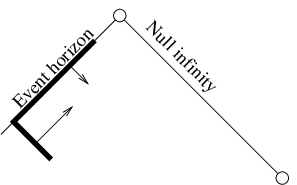} 
\]
One might expect that the smallness of the flux on the ``good'' segments
of the event horizon, which we hope to preserve,
is completely dwarfed by initial data on the conjugate null segment
depicted above, on which we know nothing. The remarkable fact is that
this is not the case. Herein lies the significance of the red-shift effect (C).

Feature (C) is tied to the very name of black holes,
as it is the red-shift effect that makes the black holes
of gravitational collapse look ``black'', rather than like frozen stars.
To measure quantitatively this effect, one considers the ratio 
of the proper time between the reception of two signals,
as measured by an observer travelling to timelike infinity,
to the proper time between the emission of the two signals by
an observer entering the black hole. As the emitter approaches
the black hole, and the proper time between his emissions tends to zero,
this ratio goes to $\infty$. In the context of the scalar field, 
this shifts the frequency of the radiation on the incoming null
ray to the red. This will allow us to 
essentially ``forget'' about this part of the initial data when
estimating as in the previous diagram.

Finally, it is worth commenting on how the power $-3$ appears. 
As noted in the above discussion, we derive decay in $v$ ultimately
from decay in $r$, specifically, from the decay in $r$ of the
quantity $\partial_v(r\phi)$, which satisfies:
\[
\partial_u(\partial_v(r\phi))=
\frac{\phi}{r^2}\frac{\partial_v r\partial_u r}
{1-\frac{2m}r}\left(m-\frac{e^2}{2r}\right)
\]
Recall that in Minkowski space,
$\partial_v(r\phi)$ vanishes (for large $v$) for spherically symmetric
solutions of the wave equation with
compactly supported data. In our case, however, due to 
backscattering off the curved background, $\partial_v(r\phi)$ will immediately
acquire an $r^{-3}$ tail on an outgoing null cone. It is for this reason
that we cannot improve beyond $-3$.

\section{A complete ``relatistic'' spherically symmetric picture}

The Einstein-Maxwell-neutral scalar field system may indeed capture
some of the phenomenology of vacuum collapse in the black hole interior.
As remarked earlier, however, it cannot 
represent  a truly complete model in spherical symmetry,
as, if $e\ne0$, complete initial data must possess two asymptotically flat
ends. Black holes are thus ``built in'' to the problem from the start.
A more interesting model arises when the matter is endowed with charge,
for instance a complex-valued scalar field $\phi$,
interacting with an electromagnetic potential $A_\mu$, defined up to a
phase (see~\cite{he:lssst}).
Large time existence results for small data
solutions of such systems have been proven by Chae~\cite{doc:ge}.
In view of the results described here, a working conjecture 
for the evolution of generic spacelike initial data
is as follows:
\begin{conjecture}
For generic, asymptotically flat, spacelike spherically symmetric
data with one end, the maximal development has 
Penrose diagram as depicted below:
\[
\includegraphics{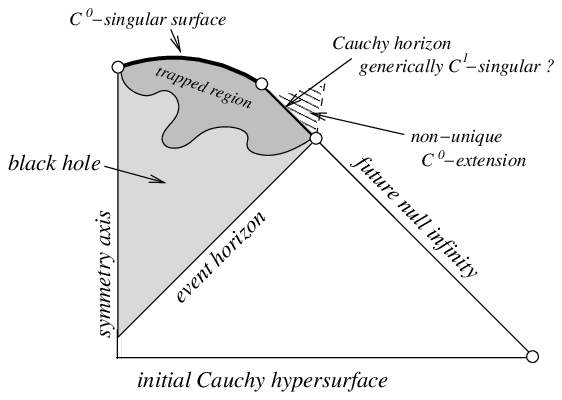}
\]
In particular, according to the above, the trapped surfaces conjecture
(and thus W.C.C.) is true, but S.C.C., as formulated,
is false.\footnote{A $C^1$-formulation, howerver,
of S.C.C. may still be true.}
\end{conjecture}

It should be emphasized once again that the primary
motivation for the matter described here 
is as a spherically symmetric
model problem for the \emph{non}-spherically symmetric \emph{vacuum}. 
Another source for interesting matter models in gravitational collapse
is kinetic theory. The reader should consult~\cite{ha:ev, rrs:reg, rr:ge}. 
In this
context, certain results have been recently extended to the charged 
case in~\cite{nnr:eid, nn:lec}, motivated by similar considerations
to those described here.

\end{document}